\documentclass[prl,superscriptaddress,floatfix, reprint,aps]{revtex4-2}

\usepackage{graphicx}
\usepackage{dcolumn}
\usepackage{bm}
\usepackage{amsmath}
\usepackage{MnSymbol}
\usepackage{wasysym}
\usepackage{xspace}
\usepackage{gensymb}
\usepackage{textcomp}
\usepackage{xcolor}
\usepackage{mathtools}

\begin{document}

\title{Confinement reduces surface accumulation of swimming bacteria}

\author{Da Wei}
\thanks{These authors contributed equally to this work}
\affiliation{Beijing National Laboratory for Condensed Matter Physics, Institute of Physics, Chinese Academy of Sciences, Beijing 100190, China}

\author{Shiyuan Hu}
\thanks{These authors contributed equally to this work}
\affiliation{School of Physics, Beihang Univerity, Beijing 100191 China}
\affiliation{Institute of Theoretical Physics, Chinese Academy of Sciences, Beijing 100190, China}

\author{Tangmiao Tang}
\affiliation{Beijing National Laboratory for Condensed Matter Physics, Institute of Physics, Chinese Academy of Sciences, Beijing 100190, China}
\affiliation{School of Physical Sciences, University of Chinese Academy of Sciences, 19A Yuquan Road, Beijing 100049, China}

\author{Yaochen Yang}
\affiliation{Institute of Theoretical Physics, Chinese Academy of Sciences, Beijing 100190, China}
\affiliation{School of Physical Sciences, University of Chinese Academy of Sciences, 19A Yuquan Road, Beijing 100049, China}

\author{Fanlong Meng}
\email{fanlong.meng@itp.ac.cn}
\affiliation{Institute of Theoretical Physics, Chinese Academy of Sciences, Beijing 100190, China}
\affiliation{School of Physical Sciences, University of Chinese Academy of Sciences, 19A Yuquan Road, Beijing 100049, China}
\affiliation{Wenzhou Institute, University of Chinese Academy of Sciences, Wenzhou, Zhejiang 325000, China}

\author{Yi Peng}
\email{pengy@iphy.ac.cn}
\affiliation{Beijing National Laboratory for Condensed Matter Physics, Institute of Physics, Chinese Academy of Sciences, Beijing 100190, China}
\affiliation{School of Physical Sciences, University of Chinese Academy of Sciences, 19A Yuquan Road, Beijing 100049, China}


\begin{abstract}
Many swimming bacteria naturally inhabit confined environments, yet how confinement influences their swimming behaviors remains unclear. Here, we combine experiments, continuum modeling and particle-based simulations to investigate near-surface bacterial swimming in dilute suspensions under varying confinement. Confinement reduces near-surface accumulation and facilitates bacterial escape. These effects are quantitatively captured by models incorporating the force quadrupole, a higher-order hydrodynamic singularity, that generates a rotational flow reorienting bacteria away from surfaces. Under strong confinement, bacterial trajectories straighten due to the balancing torques exerted by opposing surfaces. These findings highlight the role of hydrodynamic quadrupole interactions in near-surface bacterial motility, with implications for microbial ecology, infection control, and industrial applications.\\
\end{abstract}
\maketitle

Bacterial motility near surfaces is crucial for various microbial processes, including colony growth, biofilm formation, and pathogenic infections.
Over the past decades, near-surface swimming behaviors have been extensively studied, primarily in semi-infinite spaces bounded by a single surface~\cite{Frymier1995,Lauga2006,Berke2008,Li2009,Shum2010,Drescher2011,Spagnolie2012,Molaei2014,Sipos2015,Bianchi2017,Perez2019,Junot2022,Leishangthem2024,Cao2022}. Flagellated bacteria accumulate near surfaces due to the interplay of hydrodynamic interactions (HIs) and anisotropic steric interactions. A pioneering study attributed surface accumulation to the leading-order, long-range dipolar flow field~\cite{Berke2008}, while later research highlighted the essential role of direct collisions with the surface~\cite{Li2009, Bianchi2017, Bianchi2019}. Moreover, near-field HIs dictate a steady pitching angle that stabilizes surface-adjacent swimming~\cite{Spagnolie2012,Sipos2015,Leishangthem2024}, also contributing to accumulation. On the other hand, bacteria escape surface entrapment through angular diffusion~\cite{Li2009} and effective tumbling~\cite{Junot2022}.

While these studies have advanced our understanding of bacterial motility near single surfaces, the accumulation behavior and its governing mechanisms in confined geometries remain less explored. This knowledge gap is significant, as many bacteria inhabit confined spaces in both natural and clinical environments, such as sediment layers~\cite{Hewson2007}, urinary tracts~\cite{Flores2015}, and tissue interstices~\cite{Kim2010}. Boundary element simulations predict that bacteria preferentially swim along the midplane between two parallel plates when the separation falls below a critical threshold~\cite{Shum2015}. Bacterial tracking in microfluidic tunnels reveals stable swimming along the central axis in narrow tunnels~\cite{Vizsnyiczai2020}. However, experiments across various microfluidic channel designs demonstrate a complex response of bacterial motion to confinement, influenced by both bacterial and channel geometry~\cite{Tokarova2021}. 

At the microscale, confinement significantly alters both HIs and steric effects~\cite{Sano1980,Jeanneret2019,Ghosh2022}, reshaping bacterial surface entrapment. When a swimming bacterium is modeled as a collection of flow singularities~\cite{Chwang1975,Spagnolie2012}, higher-order terms, which are often negligible in unbounded fluids, can become significant under strong confinement. The influence of different singularities on bacterial distribution under varying confinement remains unclear, yet understanding this effect could inform microfluidic designs for controlling microswimmer motility~\cite{Duan15,Bechinger16,Brosseau19,Lu2023,Chen2023}.

In this letter, we combine experiments and models to investigate bacterial accumulation between two parallel plates with varying separations. As the plate separation decreases, bacterial accumulation near the surfaces reduces and can even shift into the bulk. Single bacterium tracking reveals that confinement enhances bacterial escape from surface entrapment. Simulations incorporating both HIs and steric interactions demonstrate that a higher-order singularity---the image force quadrupole---is essential to quantitatively reproduce the density profile near surfaces. This quadrupolar term induces a rotational flow, reorienting bacteria away from surface, consistent with experimental observation. While the quadrupole flow decays rapidly with distance from the surface, its rotational effect, coupled with bacterial swimming, affects population deep in the bulk even at large plate separation. In strongly confined environments, bacteria follow straighter trajectories rather than circular paths near a single surface, consistent with boundary element simulations.

\begin{figure}
\centering
\includegraphics[width=0.98\linewidth]{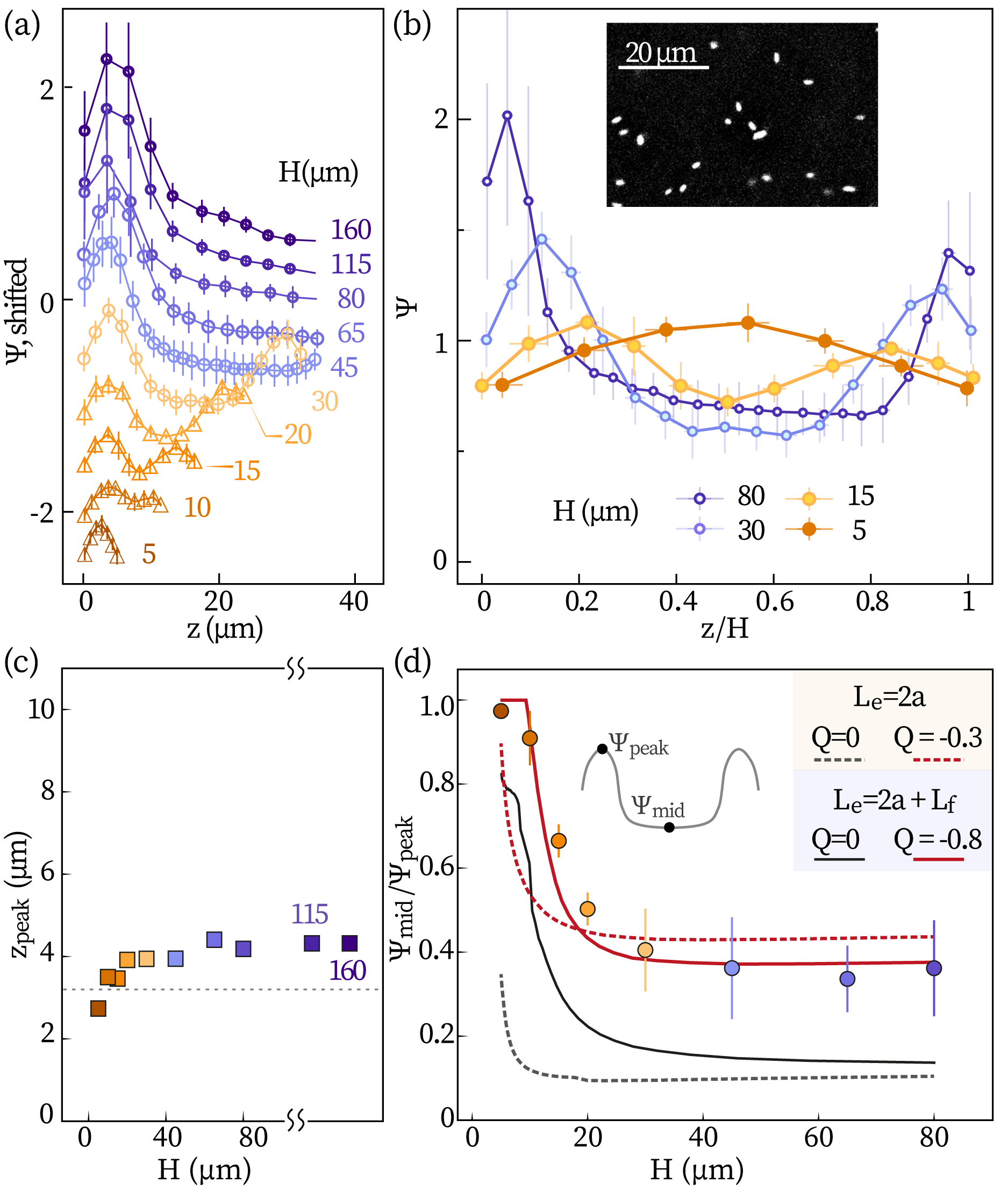}
\caption{Experimental measurements of bacterial density profiles. (a) Probability distribution function (PDF) $\Psi (z)$ for various confinement height $H$. The PDFs are vertically shifted for clarity. (b) $\Psi (z)$ plotted against scaled height $z/H$. Inset: contrast-enhanced confocal image near the bottom plate; bright regions indicate cell bodies. (c) Peak location of $\Psi(z)$, $z_{\rm peak}$ and (d) ratio $\Psi_{\rm mid}/\Psi_{\rm peak} $ as functions of $H$. Symbols and error bars denote mean $\pm$ SD over 3--4 experiments. Lines in (d) show predictions of the Smoluchowski model with fixed dipole strength $D=0.6$ pN$\cdot \mu$m, varying exclusion length $L_{\rm e}$, and quadrupole strength $Q$ (in pN$\cdot \mu$m$^2$).}
\label{fig:SA_exp}
\end{figure}

\textit{Confinement reduces surface accumulation}---We employ \emph{E. coli} as our model bacteria, composed of a 3 $\mu$m-long rod-shaped body and 10 $\mu$m-long flagellar bundle. The bacteria exhibit wild-type run-and-tumble behaviors and express green fluorescent protein. Cell concentrations ranges from 0.1 to 1 $n_0$, where $n_0 = 8 \times 10^8$~mL$^{-1}$. Bacterial suspensions are loaded into closed chambers formed by two horizontal parallel plates with separation $H$ ranging from 5 to 160 $\mu$m. In these chambers, bacteria swim at $v_0\approx15~\mu$m/s during measurements. We use spinning-disk confocal microscopy to image \textit{E. coli} swimming with high vertical spatial resolution (see Supplemental Material~\cite{SM} for details). For a given $H$, we measure the vertical density profile $\Psi(z)=A(z)/\int_0^H A(z') {\rm d}z'$, where $z$ is the height from the bottom plate and $A(z)$ is the area occupied by  bacterial bodies at $z$ [Figs.~\ref{fig:SA_exp}(a) and \ref{fig:SA_exp}(b)]. The profile $\Psi(z)$ varies progressively from surface accumulation in thick chambers, consistent with previous studies~\cite{Frymier1995,Lauga2006,Berke2008,Li2009,Drescher2011,Molaei2014,Bianchi2017,Perez2019}, to bulk accumulation under strong confinement.

The density peaks $z_{\mathrm{peak}}$ occur at a finite distance from the surfaces, consistent with prior observations with high resolution in $z$-direction~\cite{Vladescu2014droplet,Burriel2024} and similar to behavior in synthetic rod-like microswimmers~\cite{Takagi2014}. The nonzero distance results from bacterial rotation and interactions with the surface, with also possible contributions from cell-cell HIs~\cite{Htet2024hovering}, which lift cells away from the surface. As $H$ decreases, $z_{\mathrm{peak}}$ remains 4 $\mu$m for $H \gtrsim 20$~$\mu$m, but drops for $H \lesssim 20$~$\mu$m. The decrease primarily arises from confinement-induced suppression of bacterial rotation.

To quantify the effect of confinement on surface accumulation, we calculate the ratio of bacterial density at the mid-plane to its peak value, $\Psi_{\rm mid}/\Psi_{\rm peak}$, as a function of $H$ [Fig.~\ref{fig:SA_exp}(d)]. For $H > 40$ $\mu$m, $\Psi_{\rm mid}/\Psi_{\rm peak}$ plateaus at $\sim0.4$, indicating sustained surface accumulation as in semi-infinite systems. As $H$ deceases below 40 $\mu$m, $\Psi_{\rm mid}/\Psi_{\rm peak}$ rises sharply, reflecting reduced surface accumulation. For $H\lesssim 10$ $\mu$m, $\Psi(z)$ shows a single peak at mid-plane [Figs.~\ref{fig:SA_exp}(a) and \ref{fig:SA_exp}(b)], with $\Psi_{\rm mid}/\Psi_{\rm peak}\approx 1$, confirming the prediction from boundary element simulations~\cite{Shum2015}. 

To investigate how confinement reduces surface accumulation, we develop a continuum model that incorporates HIs and steric interactions between bacteria and confining surfaces. A multipolar representation of the flow field is employed to quantify bacterium-surface HIs~\cite{Spagnolie2012}. By modeling the bacterium as a spheroidal body with a slender rod-like flagellum (hereafter, the rod-spheroid model), we derive the strengths of the force dipole $D$ and force quadrupole $Q$~\cite{SM},
\begin{equation}\label{sing_strength}
\begin{aligned}
D &= \frac{1}{2}F_{\mathrm{b}} (2a + L_{\rm h}),\quad \text{with }F_{\mathrm{b}}=\zeta_\parallel v_0, \\
Q &= -\frac{1}{6}F_{\mathrm{b}} \left[L_{\rm h}^2 + 3aL_{\rm h} + a^2(3-e^2)\right],
\end{aligned}
\end{equation}
where $F_{\mathrm{b}}$ is the drag force on the cell body, $\zeta_\parallel$ the drag coefficient along the major axis, $a$ the semi-major axis length, $e$ the eccentricity, and $L_{\rm h}$ the hydrodynamic flagellum length. Using parameters from Ref.~\cite{Drescher2011}, we estimate $L_{\rm h}\approx 2.5$ $\mu$m, which is smaller than the geometric length of flagellar bundle $L_{\mathrm{f}}$. Thus, we estimate $D\approx 0.6$ pN$\cdot \mu$m, and $Q\approx -0.8$~pN$\cdot \mu$m$^2$. The corresponding source dipole strength is $S\approx -0.02$ pN$\cdot \mu$m$^2$~\cite{SM}. 
For bacteria with rod-shaped bodies, the source dipole $S$ is negligible compared to the force quadrupole $Q$, unlike in squirmer models~\cite{Pedley2016}. 

\begin{figure}[t]
\centering
\includegraphics[width=1.00\linewidth]{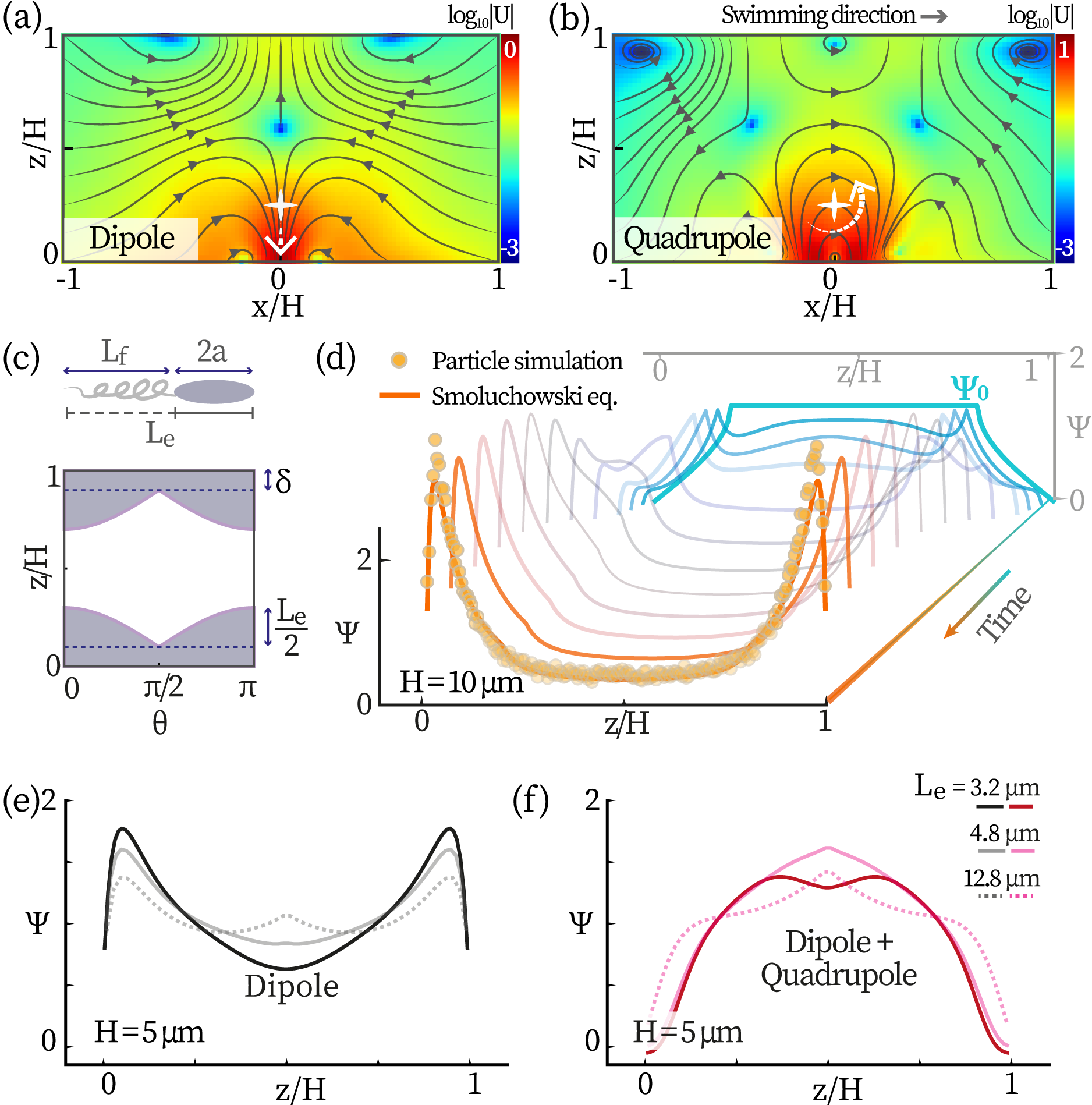}
\caption{The effect of quadrupole on bacterial population distribution. (a), (b) Image flow fields induced by a (a) force dipole and (b) force quadrupole. White crosses mark the positions of point singularities; white arrows indicate bacterial motion under image flows. (c) Model schematic of the bacterial body plan (top) and the accessible configuration space constrained by steric exclusion (bottom, white region). The exclusion length ($L_{\rm e}$) is tested to span merely the bacterial body ($2a$) and the entire bacterium ($2a+L_{\rm f}$). (d) Time evolution of $\Psi(z,t)$ from the Smoluchowski model (lines) and the particle simulations (circles), initialized with $\Psi_0$. (e) Steady-state $\Psi(z)$ for $D=0.6$ pN$\cdot \mu$m and $Q=0$. (f) Steady-state $\Psi(z)$ for $D=0.6$ pN$\cdot \mu$m and $Q=-0.5$ pN$\cdot \mu$m$^2$.}
\label{fig:theory}
\end{figure}

The surface-induced flow field at the height of the cell-body center $z$ is approximated as
\begin{equation}\label{eq_image_flow}
\mathbf{u}(z,\mathbf{p}) \approx \left[D G_{\mathrm{D}}^{\ast}(\mathbf{r},\mathbf{p}) + Q(\mathbf{p}\cdot\nabla) G_{\mathrm{D}}^{\ast}(\mathbf{r},\mathbf{p})\right]_{\mathbf{r}=z\hat{\mathbf{z}}}\cdot \mathbf{p},
\end{equation}
where the image force dipole $G^{\ast}_{\mathrm{D}}(\mathbf{r},\mathbf{p}) = (\mathbf{p}\cdot \nabla)G^{\ast}(\mathbf{r},\mathbf{r})$, with $G^{\ast}$ being the image system for a Stokeslet placed between two parallel plates, and the swimming direction $\mathbf{p} =(\cos\phi\sin\theta,\sin\phi\sin\theta,\cos\theta)$. We construct $\mathbf{u}(z,\mathbf{p})$ numerically using the exact solution of $G^{\ast}$ given in Ref.~\cite{Liron1976}, which enforces the no-slip condition on both plates when superposed with the free-space Stokeslet. Figures~\ref{fig:theory}(a) and (b) show the image flow fields of the force dipole and quadrupole, respectively, for bacteria swimming parallel to the plates. The dipolar flow field generates a drift that pulls the bacterium toward the nearest surface. In contrast, the quadrupolar flow induces no drift toward the surface but a nonzero vorticity that reorients the bacterium to swim away from the surface.

We then model bacteria as elongated spheroidal particles with major axis $L_{\rm e}$ and calculate the density profile by solving the Smoluchowski equation. The probability density function $\Psi(z,\mathbf{p},t)$ of finding a particle with height $z$ and orientation $\mathbf{p}$ at time $t$ is governed by~\cite{Doi1986,Saintillan2008}

\begin{equation}\label{continuum}
\partial \Psi/\partial t + \partial(v_z \Psi )/\partial z + \nabla_\mathbf{p}\cdot (\dot{\mathbf{p}} \Psi) = 0,
\end{equation}
where $\nabla_\mathbf{p}$ is the gradient operator on a unit sphere. The flux velocities $v_z$ and $\dot{\mathbf{p}}$ include swimming, image flow $\mathbf{u}(z,\mathbf{p})$ [Eq.~(\ref{eq_image_flow})], and thermal diffusion: $v_z = [v_0 \mathbf{p} + \mathbf{u}(z, \mathbf{p}) - \mathbf{D}_t \partial\ln\Psi/\partial z]\cdot\hat{\mathbf{z}}$ and $\dot{\mathbf{p}} = \boldsymbol{\Omega} \times \mathbf{p} - D_{\mathrm{r}} \nabla_{\mathbf{p}}\ln \Psi$, where $\mathbf{D}_t$ is the matrix of translational diffusion coefficients, $D_r$ is the rotational diffusion coefficient around the short axis, and $\boldsymbol{\Omega}$ the angular velocity given by Jeffery's equation. Both $\mathbf{D}_t$ and $D_r$ are assumed to be independent of $z$. 

Steric exclusion enforces a geometric constraint by preventing particles from penetrating the solid plates~\cite{Ezhilan2015}. Near the bottom plate, the allowed range of $\theta$ is given by
\begin{equation}\label{theta_range}
\cos^{-1}[2(z-\delta)/L_{\rm e}] \leq \theta \leq \cos^{-1}[-2(z-\delta)/L_{\rm e}],
\end{equation}
for $\delta \leq z \leq L_{\rm e}/2+\delta$ [Fig.~\ref{fig:theory}(c)]. Here, $\delta$ denotes the minimum distance between the particle and the surface, typically on the order of the particle's semi-minor axis. A particle with center close to the plate can only orient in parallel to it. A similar constraint is applied at the top plate; otherwise, $0\leq\theta\leq\pi$. Steric exclusion is imposed by setting the probability flux normal to the boundaries defined in Eq.~(\ref{theta_range}) to zero. Because the allowed range of $\theta$ depends on $z$, integrating a uniform density $\Psi_0(z,\mathbf{p})$ over $\mathbf{p}$ yields a nonuniform profile $\Psi_0(z)$, exhibiting depletion layers near the plates. This baseline distribution represents the equilibrium population distribution of nonmotile cells.

We solve Eq.~(\ref{continuum}) numerically using a finite-volume method~\cite{Ezhilan2015,SM}. Figure~\ref{fig:theory}(d) shows the evolution of the marginal distribution $\Psi(z,t)$ for $D=0.6$ pN$\cdot\mu$m and $H=10$ $\mu$m, initialized from $\Psi_0(z,\mathbf{p})$. Despite accumulation near surfaces, the depletion layers persist over time, consistence with experimental observations [Figs.~\ref{fig:SA_exp}(a) and \ref{fig:SA_exp}(b)]. Notably, steric exclusion and force dipole are insufficient to generate a central density peak, as shown by the distributions for various body length [Figs.~\ref{fig:theory}(e)]. Bulk accumulation under strong confinement emerges only when the quadrupole is included [Fig.~\ref{fig:theory}(f)]. 

We compare the numerical solutions of Eq.~(\ref{continuum}) with experiments to investigate the roles of interactions in surface accumulation under confinement [Fig.~\ref{fig:SA_exp}(d)]. For $D = 0.6$ pN$\cdot\mu$m and $Q = 0$, simulated $\Psi_{\rm mid}/\Psi_{\rm peak}$ remains significantly below experimental measurements across all $H$. Increasing $Q$ rises $\Psi_{\rm mid}/\Psi_{\rm peak}$. Steric exclusion contributes to the rapid increase in $\Psi_{\rm mid}/\Psi_{\rm peak}$ for $H \lesssim L_{\rm e}$. Using the bacterial body length $2a = 3.2$ $\mu$m as $L_{\rm e}$, $Q \approx -0.3$ pN$\cdot\mu$m$^2$ fits the experimental data at large $H$, but exhibits an offset at small $H$. With $L_{\rm e}=2a + L_{\rm f} = 12.8$ $\mu$m, approximately the full bacterium length, and using the estimated quadrupole $Q \approx -0.8$ pN$\cdot\mu$m$^2$ [Eq.~(\ref{sing_strength})], the experimental data is accurately reproduced across all $H$ without introducing fitting parameters. Comparisons of the full bacterial distributions between experiments and models are provided in the Supplemental Material~\cite{SM}. These results highlight the essential roles of force quadrupole and flagellar steric interaction in shaping the density profile.

\begin{figure}
\centering
\includegraphics[width=1.00\linewidth]{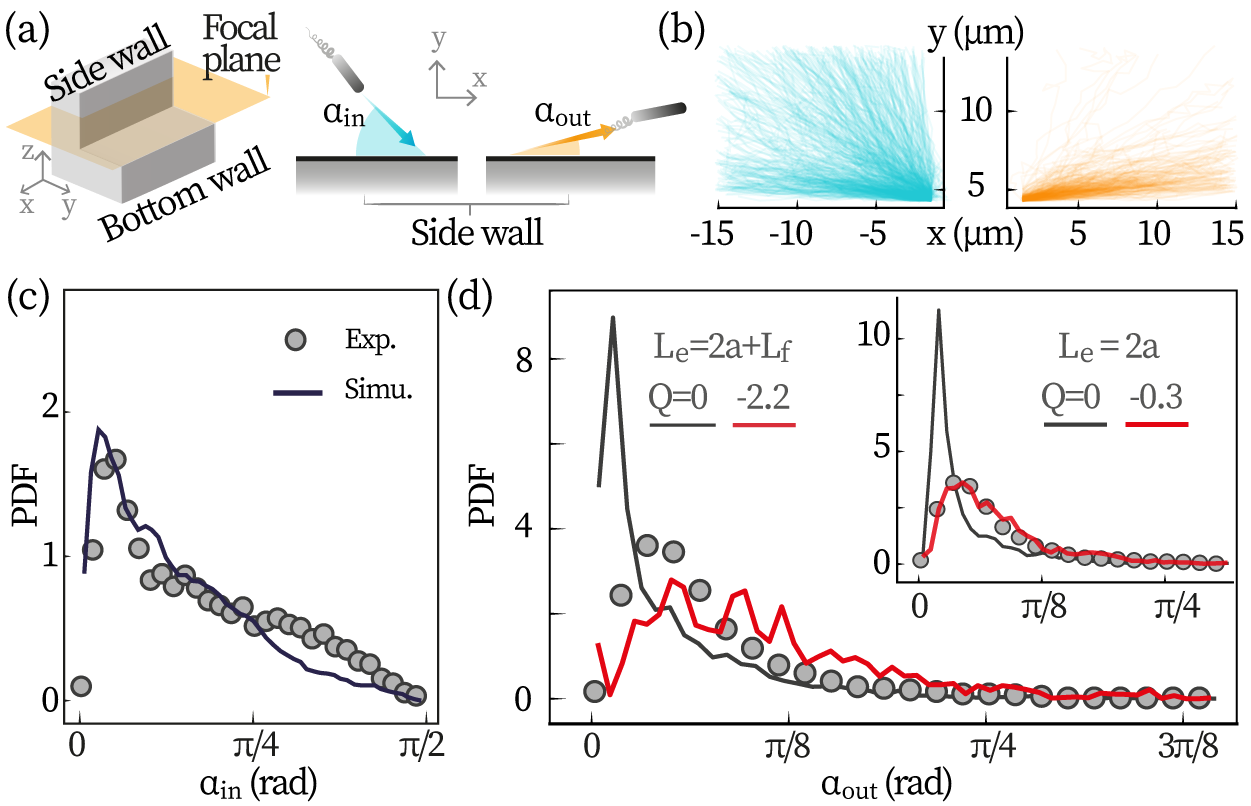}
\caption{Side-wall collision dynamics and the role of force quadrupole. (a) Left: schematic of the side-wall experiment. Right: definition of incident and outgoing angles, $\alpha_{\rm in}$ and $\alpha_{\rm out}$. (b) Experimentally measured incident and outgoing trajectories. Each trajectory is translated such that the closest approach to the wall aligns with the origin. (c) Distributions of $\alpha_{\rm in}$ measured experimentally (circles) and imposed in particle-based simulations (line). (d) Distribution of $\alpha_{\rm out}$ from experiment (circles) and simulations (line) with dipole strength $D=1.2$ pN$\cdot \mu$m, varying bacterial exclusion length $L_{\rm e}$, and quadrupole strength $Q$ (in pN$\cdot \mu$m$^2$). The value $Q=-2.2$ pN$\cdot \mu$m$^2$ is calculated for an elongated bacterium from the rod-spheroid model, rather than a fitting parameter. Inset: $Q=-0.3$ pN$\cdot \mu$m$^2$ gives the best fit for $L_\mathrm{e}=2a$. Simulations use a rotational diffusion coefficient $D_r=0.06$ rad$^2$/s, consistent with Ref.~\cite{Drescher2011}.}
\label{fig:sidewall}
\end{figure}
\textit{Quadrupole-enhanced bacterium escape}---To elucidate the role of the quadrupole in bacterium-surface interaction, we track swimming bacteria in a horizontal plane as they collide with a vertical wall [Fig.~\ref{fig:sidewall}(a)]. The observation plane is positioned more than 30 $\mu$m away from the top and bottom surfaces, where the influence of both plates is negligible. 
To enhance tracking accuracy, we use cephalexin-treated cells with elongated cell bodies $2a \approx 6.0$ $\mu$m rather than wild-type cells~\cite{Wei2024}. We identify incident events, in which bacteria approach and become trapped near the wall, and outgoing events, in which they escape from the entrapped region. As most trajectories do not include both incident and outgoing events, we analyze their angular distributions separately~\cite{SM}. Figure~\ref{fig:sidewall}(b) shows the representative tracks, whose closest approaching points to the wall are shifted and aligned at the origin. Clearly, the outgoing angle $\alpha_{\rm out}$ exhibits a narrower distribution than the incident angle $\alpha_{\rm in}$.

To resolve the collision dynamics with the wall, we simulate bacteria as active Brownian particles~\cite{SM}. Our simulations validate that, in the two-plate setup, the steady-state distribution $\Psi(z)$ from particle simulations agrees with the continuum model [Fig.~\ref{fig:theory}(d)]. We then simulate side-wall collisions by constructing the image flow field [Eq.~(\ref{eq_image_flow})] using the Blake tensor~\cite{Blake1971}. For elongated bacteria, we estimate the dipole and quadrupole strengths using the rod-spheroid model [Eq.~(\ref{sing_strength})] as $D\approx 1.2$ pN$\cdot\mu$m and $Q\approx -2.2$ pN$\cdot\mu$m$^2$, respectively. In simulations, particles are initialized near the surface, with incident angles $\alpha_{\rm in}$ sampled from the experimental distribution [Fig.~\ref{fig:sidewall}(c)]. When the quadrupole is neglected ($Q=0$), the distribution of $\alpha_{\rm out}$ displays a sharp peak at a smaller angle than observed experimentally [Fig.~\ref{fig:sidewall}(d)]. In contrast, incorporating the quadrupole term ($Q=-2.2$ pN$\cdot\mu$m$^2$) flattens the distribution of $\alpha_{\rm out}$ that matches the experimental data without fitting parameters. The minor deviation may arise from the flexible joint between the flagella and the cell body~\cite{Nord2022}, which reduces the effective bacterial length $L_\mathrm{e}$. Using the lower bound $L_\mathrm{e}=2a$, we find that a nontrivial quadrupole term is required to fit the experimental data.

In addition to confocal imaging, we use defocused fluorescent microscopy to track 3D bacterial motion~\cite{Wu2005}. By focusing on the top surface of the chamber, out-of-focus bacteria appear as diffraction rings. The ring radius is proportional to the distance between the bacterium and the focal plane. An escape event is defined as a near-surface bacterium swimming more than 5 $\mu$m away from the surface. The surface escape rate is the number of escape events per cell per unit time~\cite{Cao2022,SM}. In an 80 $\mu$m-thick chamber, we measure an escape rate of 0.038 $\pm$ 0.003 s$^{-1}$, corresponding to a mean trapping time $\tau \approx 26$ s, consistent with the previously reported value of 21 s~\cite{Junot2022}. As $H$ decreases to 30 $\mu$m, the escape rate increases by 46$\%$, indicating enhanced bacterial escape from surface entrapment under confinement. Simulations incorporating a force quadrupole predict a $\approx$ 50$\%$ increase in escape rate as $H$ decreases from 80 $\mu$m to 30 $\mu$m, compared to $<$ 30 $\%$ in simulations neglecting the quadrupole. These results demonstrate that the force quadrupole enhances bacterial escape from surfaces, thereby reducing surface accumulation.

\begin{figure}
\centering
\includegraphics[width=1.00\linewidth]{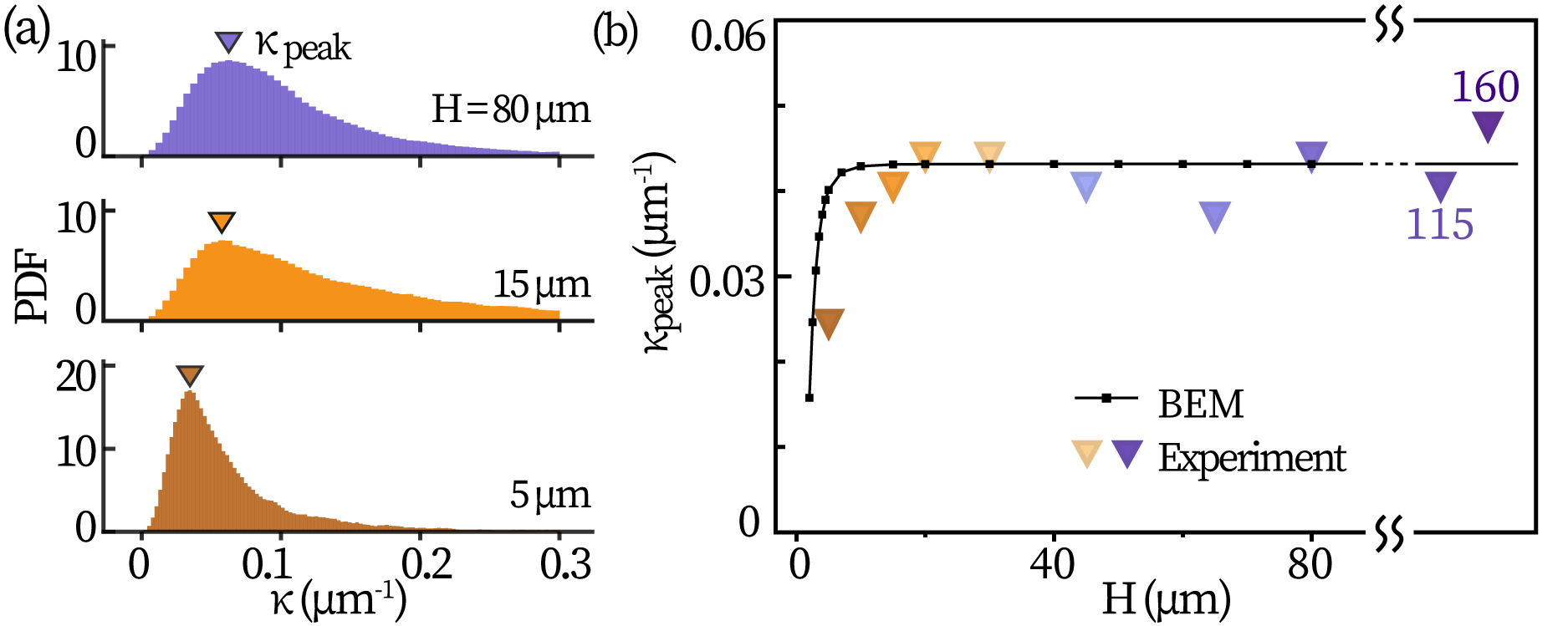}
\caption{Curvature of bacterial swimming trajectories under confinement. (a)  Probability distributions of trajectory curvature $\kappa$ measured experimentally for various $H$. Inverted triangles mark the most probable curvatures $\kappa_{\mathrm{peak}}$. (b) $\kappa_{\mathrm{peak}}$ as a functions of $H$. Solid line shows prediction from BEM simulations.}
\label{fig:curvature}
\end{figure}

\textit{Bacterial trajectories}---Flagellated bacteria are known to swim in circles near solid surfaces, driven by a surface-induced hydrodynamic torque acting on the cells~\cite{Lauga2006}. To probe the effect of confinement on circular swimming, we use confocal microscopy to track two-dimensional bacterial trajectories near the bottom plate in chambers of varying height $H$. Circular motion is characterized by the trajectory curvature $\kappa$. The probability distribution of $\kappa$ peaks near 0.042 $\mu$m$^{-1}$ and displays a long tail for $H > 15$ $\mu$m [Fig.~\ref{fig:curvature}(a)]. As $H$ decreases below 15 $\mu$m, the peak position $\kappa_{\rm peak}$ shifts to lower values and the tail shortens, indicating straighter trajectories under confinement. 

We perform numerical simulations using the boundary element method (BEM)~\cite{SM,Liron1976,Smith2009_2}, yielding results in quantitative agreement with experimental observations [Fig.~\ref{fig:curvature}(b)]. Near a surface, circular motion arises from an effective hydrodynamic torque generated by interactions between the surface and the rotating flagella and cell body~\cite{Lauga2006}. In confined environments, the opposing surfaces exert hydrodynamic torques in opposite directions, canceling each other~\cite{Liron1976}. As a result, under strong confinement, both the net torque and trajectory curvature decrease with decreasing $H$. The circular swimming arises to leading order from the image flow of the rotlet dipole. Since near a single plate $\kappa \sim 1/z^4$~\cite{Lauga2020}, the top plate starts to affect $\kappa$ only when it is sufficiently close to the bacteria, resulting in a sharp variation of $\kappa$ at small $H$ [Fig.~\ref{fig:curvature}(b)].

\textit{Discussion}---Combining experiments, continuum theory and particle-based simulations, we show that swimming bacteria tend to
escape surface entrapment and accumulate near the mid-plane in confined geometries,
which originates from fluid flows induced by force quadrupole. The force quadrupole introduces a new mechanism facilitating detachment from surfaces, which can act in parallel with the tumble-mediated escape~\cite{Junot2022}. These findings advance the understanding of microswimmer surface accumulation under confinement and highlight the fundamental role of force-quadrupole hydrodynamics, often neglected in prior studies. 

We use spheroidal particles to approximate the bacterial dynamics in the image flow field. Despite its simplicity, this model quantitatively reproduces experimental observations merely using parameters derived from bacterial geometry and motility. The strengths of hydrodynamic singularities can be tuned by altering bacterial geometries and motility, suggesting a potential route to control microswimmer distributions in confinement~\cite{Shum2010,Spagnolie2012,Tokarova2021}.

\bigskip

\begin{acknowledgments}
We acknowledge supports from the National Natural Science Foundation
of China (No. T2221001, 12474206, 12204525, 12247130, 12275332 and 12047503), and the Fundamental Research Funds for the Central Universities (Beihang University), Wenzhou Institute (No. WIUCASQD2023009), and Beijing National Laboratory for Condensed Matter Physics (No. 2023BNLCMPKF005).
\end{acknowledgments}

\bibliographystyle{apsrev4-2}
\bibliography{reference}

\end{document}


\title{Supplemental Material for \\
Confinement reduces surface accumulation of swimming bacteria}

\author{Da Wei}
\thanks{These authors contributed equally to this work}
\affiliation{Beijing National Laboratory for Condensed Matter Physics, Institute of Physics, Chinese Academy of Sciences, Beijing 100190, China}

\author{Shiyuan Hu}
\thanks{These authors contributed equally to this work}
\affiliation{School of Physics, Beihang Univerity, Beijing 100191 China}
\affiliation{Institute of Theoretical Physics, Chinese Academy of Sciences, Beijing 100190, China}

\author{Tangmiao Tang}
\affiliation{Beijing National Laboratory for Condensed Matter Physics, Institute of Physics, Chinese Academy of Sciences, Beijing 100190, China}
\affiliation{School of Physical Sciences, University of Chinese Academy of Sciences, 19A Yuquan Road, Beijing 100049, China}

\author{Yaochen Yang}
\affiliation{Institute of Theoretical Physics, Chinese Academy of Sciences, Beijing 100190, China}
\affiliation{School of Physical Sciences, University of Chinese Academy of Sciences, 19A Yuquan Road, Beijing 100049, China}

\author{Fanlong Meng}
\email{fanlong.meng@itp.ac.cn}
\affiliation{Institute of Theoretical Physics, Chinese Academy of Sciences, Beijing 100190, China}
\affiliation{School of Physical Sciences, University of Chinese Academy of Sciences, 19A Yuquan Road, Beijing 100049, China}
\affiliation{Wenzhou Institute, University of Chinese Academy of Sciences, Wenzhou, Zhejiang 325000, China}

\author{Yi Peng}
\email{pengy@iphy.ac.cn}
\affiliation{Beijing National Laboratory for Condensed Matter Physics, Institute of Physics, Chinese Academy of Sciences, Beijing 100190, China}
\affiliation{School of Physical Sciences, University of Chinese Academy of Sciences, 19A Yuquan Road, Beijing 100049, China}

\maketitle

\section{Materials and methods}\label{secMeth}

\textit{Bacteria culture and sample preparation.}
Wild-type \textit{E. coli} (BW25113) expressing green fluorescent protein is employed in this work~\cite{Wei2024}. Bacteria are first cultured overnight at 37.0~°C in terrific broth (TB) medium [trypotone 1.2$\%$ (w/v), yeast extract 2.4$\%$ (w/v) and glycerol 0.4$\%$ (v/v)]  under shaking at 250 rpm. The saturated culture is diluted 100 fold in fresh TB medium and incubated at 30°C under identical shaking conditions.  After 6.5 hours, bacteria are harvested by gentle centrifugation [800 $\times$ g, 5 min], washed, and resuspended in motility buffer [0.01 M potassium phosphate, 0.067 M NaCl, and 10 M EDTA, pH 7.0]. The suspension is adjusted to working concentrations between $0.1~n_0$ and $1~n_0$, where $n_0 = 8 \times 10^8$~mL$^{-1}$. The relative density profile remain unchanged across this concentration range. To induce cell elongation, cephalexin ($\sim$30~\textmu{g}/ml) is added during the second culture. The resulting cells exhibit an average body length of $6.0 \pm 1.3~\mu$m (mean $\pm$ SD). Experimental chambers are assembled using glass slides and coverslips, spaced with double-sided tape of varying thickness. Chambers typically measure 18 mm $\times$ 5 mm. Chambers are filled with bacteria suspensions and sealed with UV-curable adhesive (NOA81).

\textit{Video microscopy and image analysis.}
Spinning-disk confocal microscopy is used to track bacterial swimming in 2D at varying distances from the bottom plates. Imaging is performed on a Nikon Ti2-E 
microscope equipped with a Yokugawa CSU-X1 unit and through 40$\times$ / 20$\times$ objectives. Videos are recorded at 20 frames per second for 30~s using a scientific complementary metal-oxide semiconductor (sCMOS) camera. To observe swimming bacteria across the confined chamber, the focal plane is adjusted in 1~\um steps along the vertical axis $z$. A 2D band-pass filter is applied to suppress background intensity variations and remove noise. Images are adjusted via histogram saturation and binarized prior to cell-body detection. Cell detection and tracking are performed using TrackMate plugin in ImageJ~\cite{Tinevez2017Trackmate}.
A custom Python script is used to extract the fraction of occupied pixels and compute bacterial density. 
Bacterial trajectories are selected for curvature analysis based on three criteria: (1) trajectory duration exceeds 0.5 s; (2) mean speed exceeds 4~\ums to exclude immotile cells; (3) aspect ratio exceeds 3, ensuring in-plane swimming.

For side-wall interaction experiments, the focal plane is fixed 30~\um above the bottom surface. We analyze the angular distributions of incident and outgoing trajectories separately for two main reasons. (1) If a cell's incident trajectory lies within the focal plane ($xy$-plane), its outgoing trajectory typically does not remain in that plane, and vice versa.
(2) After incidence, bacteria often become entrapped and swim parallel to the wall for some distance. During this phase, cells may move out of focus or leave the field of view. Most importantly, they become indistinguishable from other wall-trapped cells.
As a result, obtaining full trajectories---including incidence, entrapment, and outgoing segments for the same cell---is challenging. Therefore, we measure angular distributions for incidence and outgoing events separately. In practice, these angles are determined by linear least-squares fitting of trajectory segments located at least 4~\um from the side wall. We first discard near-wall points (y $\leq$ 4~\um) from all trajectories,after which trajectories are typically divided into segments. Segments longer than 0.5 s ($\approx$ 10 points) are categorized as incident or outgoing, depending on whether their mean velocity is directed toward or away from the wall. Finally, for each incident or outgoing segment, the angle is determined by linear fitting.

Defocused fluorescence microscopy is employed to track 3D trajectories of individual cells. Imaging is performed using a Nikon TI2-E microscope with a 60× objective. Fluorescence images are acquired in epifluorescence mode using a DAPI filter set and  a mercury pre-centered fiber illuminator. Recordings are acquired at 10 frames per second using an sCMOS camera. The field of view is $\pi$ (160~$\mu$m)$^2$. 3D trajectories are reconstructed from recorded videos using ImageJ and custom Python scripts. The focal plane is adjusted image immotile bacteria adhered to the top surface. Bacterial bodies in focus appear as bright spots, out-of-focus cells form rings. The ring radius is proportional to the bacterium's distance from the focal plane. This relationship is calibrated by varying the focal plane and measuring the corresponding ring radius. During calibration, the focal plane is set to the first layer of swimming bacteria. Bacteria are classified as near-surface swimmers when they remain within $\approx 5 \mu$m of the surface for $>$ 1 s. An escaping event is defined when a near-surface swimmer move farther than 5 $\mu$m from the surface.


\section{Hydrodynamic singularities of a model bacterium}\label{sec1}
\begin{figure}[t]
\centering
\includegraphics[bb= 0 0 365 90, scale=0.8, draft=false]{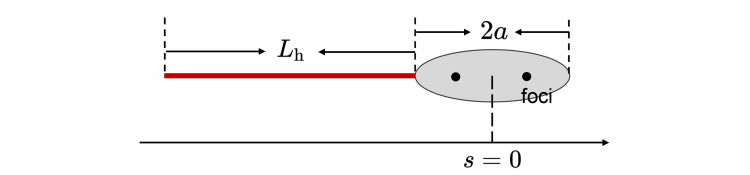}
\caption{A bacterium model consists of a rigid rod and a spheroid particle. This simplified rod-spheroid model is used to estimate the strengths of various hydrodynamic singularities.}
\label{figS1}
\end{figure}
We compute the hydrodynamic singularities of swimming  \textit{E. coli} using a minimal model bacterium. The cell body is modeled as a spheroid with major and minor axis lengths $2a$ and $2b$, respectively [Fig.~\ref{figS1}]. The aspect ratio is defined as $\beta=a/b > 1$, and the eccentricity is given by $e=\sqrt{1-1/\beta^2}$. The helical flagellum bundle is approximated by a slender rod of effective length $L_{\rm h}$. Propulsion is modeled by applying a distributed tangential force density of total magnitude $F$ along the rod. Let $\mathbf{p}$ denote the swimmer's orientation. Under the force-free condition, the drag forces exerted by the rod and the cell body along $\mathbf{p}$ are given by
\begin{equation}
F_{\mathrm{b}} = \frac{\zeta_{\parallel}}{\zeta_{\parallel}+4\pi\mu L_{\rm h} /c} F = \gamma_{\mathrm{b}} F, \quad F_{\mathrm{r}} = (1-\gamma_{\mathrm{b}}) F,
\end{equation}
where $4\pi\mu/c$ and $\zeta_{\parallel}$ are the parallel drag coefficients of the rod and the spheroid, respectively. The parameter $c=|\ln (\epsilon^2 e)|$, where $\epsilon$ is the rod aspect ratio. The ratio $\gamma_{\mathrm{b}}$ denotes the ratio of the drag coefficient of the body to the total drag coefficient.

The centerline of the whole model swimmer is parameterized by $s \in [-L_{\rm h}-a, a]$, with $s=0$ corresponding to the body center. The position of the centerline is $\mathbf{r}(s) = \mathbf{r}_{\mathrm{b}} + s \mathbf{p}$, where $\mathbf{r}_{\mathrm{b}}$ is the position of the body center. ~\citet{Chwang1975} have shown that the flow field of a translating spheroid can be represented by a distribution of Stokeslet and source dipole located between two focal points. Therefore, the velocity field generated by the model swimmer at position $\mathbf{x}$ can be expressed as
\begin{equation}\label{model_bacterium}
\begin{aligned}
8\pi\mu \mathbf{U}(\mathbf{x}) &= \int_{-L_{\rm h}-a}^{-a} \left(-\frac{F}{L_{\rm h}} + \frac{F_{\mathrm{r}}}{L_{\rm h}}\right) G[\mathbf{r}(s),\mathbf{x}]\cdot \mathbf{p}\,ds + \int_{-ae}^{ae} \frac{F_{\mathrm{b}}}{2ae} G[\mathbf{r}(s), \mathbf{x}]\cdot \mathbf{p}\,ds \\
&\quad - \int_{-ae}^{ae} \frac{F_{\mathrm{b}}}{2ae}\frac{1}{2}\left(a^2-\frac{s^2}{e^2}\right)(1-e^2) G_{\mathrm{SD}}[\mathbf{r}(s), \mathbf{x}]\cdot \mathbf{p} \,ds,
\end{aligned}
\end{equation}
where $G$ and $G_{\mathrm{SD}}$ are the Stokeslet and source dipole, respectively, 
\begin{equation}
\begin{aligned}
G(\mathbf{r},\mathbf{x}) &= \frac{\mathbf{I}}{R} + \frac{\mathbf{R}\mathbf{R}}{R^3},\quad \text{with }\mathbf{R} = \mathbf{x}-\mathbf{r}, \\
G_{\mathrm{SD}}(\mathbf{r},\mathbf{x}) &= -\frac{1}{2}\nabla^2_{\mathbf{r}}G(\mathbf{r},\mathbf{x}).
\end{aligned}
\end{equation}

We perform multipolar expansion around the spheroid center in the far-field limit, $|\mathbf{r}(s)-\mathbf{r}_{\mathrm{b}}| \ll |\mathbf{x}|$,
\begin{equation}
\begin{aligned}
\mathbf{U}(\mathbf{x}) & \approx -\frac{1}{8\pi\mu}\int_{-L_{\rm h}-a}^{-a} \frac{F_{\mathrm{b}}}{L_{\rm h}} \left[G(\mathbf{r}_{\mathrm{b}},\mathbf{x}) + s (\mathbf{p}\cdot\nabla_{\mathbf{r}})G(\mathbf{r}, \mathbf{x})|_{\mathbf{r}=\mathbf{r}_{\mathrm{b}}} + \frac{1}{2}s^2 (\mathbf{p}\cdot \nabla_{\mathbf{r}}) (\mathbf{p}\cdot \nabla_{\mathbf{r}}) G(\mathbf{r}, \mathbf{x})|_{\mathbf{r}=\mathbf{r}_{\mathrm{b}}}\right]\cdot \mathbf{p} \,ds \\
&\quad + \frac{1}{8\pi\mu}\int_{-ae}^{ae}\frac{F_{\mathrm{b}}}{2ae}\left[G(\mathbf{r}_{\mathrm{b}},\mathbf{x}) + s(\mathbf{p}\cdot\nabla_{\mathbf{r}})G(\mathbf{r},\mathbf{x})|_{\mathbf{r}=\mathbf{r}_{\mathrm{b}}} + \frac{1}{2}s^2(\mathbf{p}\cdot\nabla_{\mathbf{r}})(\mathbf{p}\cdot\nabla_{\mathbf{r}})G(\mathbf{r},\mathbf{x})|_{\mathbf{r}=\mathbf{r}_{\mathrm{b}}} \right]\cdot \mathbf{p}\,ds \\
&\quad - \frac{1}{8\pi\mu}\int_{-ae}^{ae} \frac{F_{\mathrm{b}}}{2ae}\frac{1}{2}\left(a^2 - \frac{s^2}{e^2}\right) (1-e^2)G_{\mathrm{SD}}(\mathbf{r}_{\mathrm{b}},\mathbf{x})\cdot \mathbf{p} \,ds \\
& = \frac{1}{8\pi\mu}\left[D G_{\mathrm{D}}(\mathbf{r}_{\mathrm{b}}, \mathbf{x}) + Q G_{\mathrm{Q}}(\mathbf{r}_{\mathrm{b}}, \mathbf{x}) + S G_{\mathrm{SD}}(\mathbf{r}_{\mathrm{b}}, \mathbf{x})\right]\cdot \mathbf{p},
\end{aligned}
\end{equation}
where the Stokeslet dipole and quadrupole are defined as
\begin{equation}
\begin{aligned}
G_{\mathrm{D}}(\mathbf{r}, \mathbf{x}) &= \mathbf{p}\cdot \nabla_{\mathbf{r}}G(\mathbf{r},\mathbf{x}),\\
G_{\mathrm{Q}}(\mathbf{r}, \mathbf{x}) &= (\mathbf{p}\cdot \nabla_{\mathbf{r}})(\mathbf{p}\cdot \nabla_{\mathbf{r}})G(\mathbf{r},\mathbf{x}).
\end{aligned}
\end{equation}
The strengths of these three singularities are computed as
\begin{gather}
D = -\int_{-L_{\rm h}-a}^{-a} \frac{F_{\mathrm{b}}}{L_{\rm h}} s\,ds + \int_{-ae}^{ae} \frac{F_{\mathrm{b}}}{2ae}s\, ds = \frac{1}{2}F_{\mathrm{b}} (2a + L_{\rm h}), \label{dipole} \\
Q = -\int_{-L_{\rm h}-a}^{-a} \frac{1}{2} \frac{F_{\mathrm{b}}}{L_{\rm h}} s^2\,ds + \int_{-ae}^{ae} \frac{1}{2}\frac{F_{\mathrm{b}}}{2ae} s^2 \,ds = -\frac{1}{6}F_{\mathrm{b}} \left[L_{\rm h}^2 + 3a L_{\rm h} + a^2(3-e^2)\right], \label{quadrupole} \\
S = -\int_{-ae}^{ae}\frac{F_{\mathrm{b}}}{2ae} \frac{1}{2}\left(a^2-\frac{s^2}{e^2}\right)(1-e^2)\,ds = -\frac{1}{3}F_{\mathrm{b}} a^2 (1-e^2). \label{source_dipole}
\end{gather}

In Ref.~\cite{Drescher2011}, the reported bacterial swimming speed is $v_0 \approx 20$ $\mu$m/s, with a measured dipole strength $D \approx 0.8$ pN$\cdot\mu$m. We use this measurement to estimate the effective flagellum length $L_{\rm h}$. The drag force on the cell body $F_{\mathrm{b}} = \zeta_{\parallel}v_0$. Substituting into Eq.~(\ref{dipole}), we obtain $L_{\rm h}\approx 2.5$ $\mu$m, such that for $v_0 \approx 20$ $\mu$m/s, the calculated 
 dipole strength matches the measured value.  The estimated effective flagella length $L_{\rm h}\approx 2.5$ $\mu$m, obtained via multipole expansion, is smaller than the actual flagellum length of \textit{E. coli}. This suggests that when multiple flagella form a bundle, the thrust and drag are not uniformly distributed along its axis. In our experiments, $v_0 \approx 15$ $\mu$m/s. Using the estimated $L_{\rm h}$, we find $D\approx 0.6$ pN$\cdot\mu$m and $Q\approx -0.8$ pN$\cdot\mu$m$^2$. The geometric parameters used in the model, along with the singularities estimated, are summarized in Table~\ref{table1}. Since $S \ll Q$, the contribution of the source dipole is neglected. For elongated cells, experimental results show that the swimming velocity is comparable to that of wild-type bacteria. The effective flagellum length $L_{\rm h}$ is assumed to be the same as wild type cells. The singularity strengths for elongated cells are estimated to be $D\approx 1.2$ pN$\cdot\mu$m and $Q\approx -2.2$ pN$\cdot\mu$m$^2$.
\begin{table}[t]
\centering
 \begin{tabular}{c|c|c} 
 Quantity & Symbol & Value \\ 
 \hline
 semi-major axis length & $a$ & 1.6 $\mu$m \\
 cell body aspect ratio & $\beta$ & 3 \\
 bacteria swimming velocity & $v_0$ & 15 $\mu$m/s \\
 effective flagellum length & $L_{\rm h}$ & 2.5 $\mu$m \\
 Stokeslet dipole strength & $D$ & 0.6 pN$\cdot$$\mu$m \\
 Stokeslet quadrupole strength & $Q$ & -0.8 pN$\cdot$$\mu$m$^2$ \\ 
 Source dipole strength & $S$ & -0.02 pN$\cdot$$\mu$m$^2$\\
 \hline
 \end{tabular}
 \caption{Physical parameters of the model bacterium shown in Fig.~\ref{figS1}.}
 \label{table1}
\end{table}

\section{Numerical solution of Smoluchowski equation}
We solve the Smoluchowski equation [Eq. (3) in the main text] numerically using a finite-volume method~\cite{Ezhilan2015}. We introduce the probability flux $\mathbf{J} = v_z \Psi + \dot{\mathbf{p}}\Psi = J_z \hat{\mathbf{z}} + J_{\theta}\hat{\boldsymbol{\theta}} + J_{\phi}\hat{\boldsymbol{\phi}}$. The components of the probability flux are given by
\begin{gather}\label{flux}
J_z = v_0\cos\theta\Psi + u_z \Psi - (D\cdot\hat{\mathbf{z}}) \frac{\partial \Psi}{\partial z},\\
J_{\theta} = \Omega_{\phi} \Psi - D_{\mathrm{r}} \frac{\partial \Psi}{\partial\theta}, \\
J_{\phi} = -\Omega_{\theta}\Psi - D_{\mathrm{r}} \frac{1}{\sin\theta} \frac{\partial \Psi}{\partial \phi}.
\end{gather}

\textit{Boundary condition.}---At fixed $z$, steric interactions near the walls restrict the accessible range of the polar angle $\theta$. The solution domain is bounded by four hypersurfaces:
\begin{equation}\label{theta_range}
\begin{dcases}
\cos^{-1}[(z-\delta)/a] \leq \theta \leq \cos^{-1}[-(z-\delta)/a], & \delta \leq z \leq \delta + a,\\
\cos^{-1}[-(z-H+\delta)/a] \leq \theta \leq \cos^{-1}[(z-H+\delta)/a], & H-\delta-a \leq z \leq H-\delta,
\end{dcases}
\end{equation}
otherwise, $0 \leq \theta \leq \pi$. Here, $\delta = a/2$ is the minimum distance between the particle center and the walls. To enforce no penetration through solid walls, the probability flux $\mathbf{J}$ along the normal direction $\hat{\mathbf{n}}(\theta, z)$ must vanish at the boundaries: $\mathbf{J}\cdot \hat{\mathbf{n}} = 0$. For example, in the range $\delta \leq z \leq \delta+a$, the no-flux condition leads to the following boundary conditions:
\begin{equation}\label{zero_normal_flux}
\begin{dcases}
J_z + a\sin\theta J_{\theta} = 0, & \text{for }\theta = \cos^{-1}[(z-\delta)/a]\ (<\pi/2) \\
J_z - a\sin\theta J_{\theta} = 0, & \text{for }\theta = \cos^{-1}[-(z-\delta)/a]\ (>\pi/2).
\end{dcases}
\end{equation}

\begin{figure}[t]
\centering
\includegraphics[bb= 0 0 520 380, scale=0.5, draft=false]{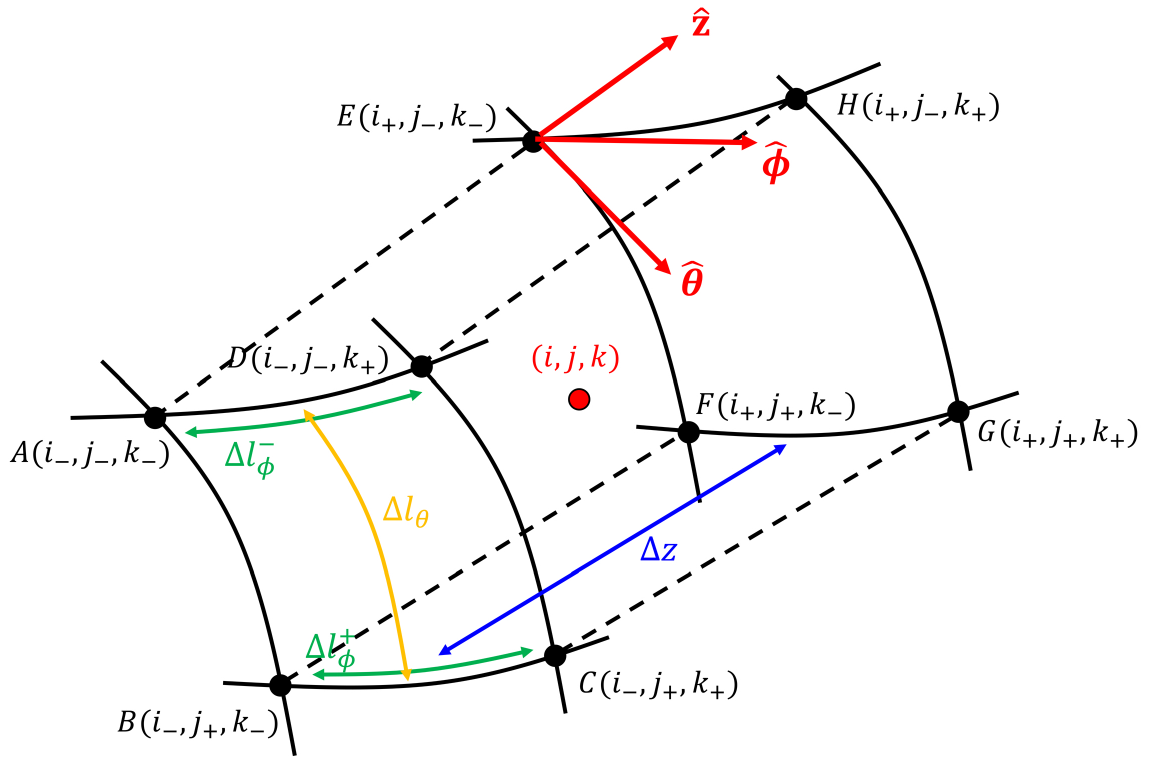}
\caption{A control volume at indices $(i,j,k)$.}
\label{figS2}
\end{figure}
\textit{Finite-volume method}.---The solution domain $(z, \theta, \phi)$ is discretized into finite volumes indexed by $(i,j,k)$. Nodal points $(i,j,k)$ are located at the centers of the corresponding finite volumes. Let $\Psi_{i,j,k}$ denote the value of $\Psi$ at the nodal point $(z_i, \tilde{\theta}_j, \phi_k)$. We define $\tilde{\theta} = \cos\theta \in [1,-1]$ for $\theta \in [0,\pi]$, and discretize $\tilde{\theta}$ rather than $\theta$. We begin by describing the computational cell, as shown in Fig.~\ref{figS2}. The nodal points are
\begin{equation}
\begin{aligned}
z_i &= \delta + \Delta z \left(i-\frac{1}{2}\right) \text{ with }\Delta z = \frac{1-2\delta}{N_z}, \quad i = 1,2,\cdots, N_z, \\
\tilde{\theta}_j  &= 1 - \Delta\tilde{\theta} \left(j - \frac{1}{2}\right)  \text{ with }\Delta \tilde{\theta} = \frac{2}{N_{\theta}}, \quad j=1,2,\cdots, N_{\theta}, \\
\phi_k &= \Delta\phi \left(k - \frac{1}{2}\right) \text{ with } \Delta\phi = \frac{2 \pi}{N_{\phi}}, \quad k = 1,2, \cdots, N_{\phi}.
\end{aligned}
\end{equation}
By definition, $\tilde{\theta}_j$ decreases with $j$. To ensure that nodal points lie exactly on the hypersurfaces, we impose the constraint:
\begin{equation}
\Delta z = a\Delta \tilde{\theta} = \frac{2 a}{N_{\theta}}, \quad N_z = \frac{(1-2\delta)N_\theta}{2 a}.
\end{equation}
The surfaces of the control volume are located at $i_{\pm} = i \pm 1/2$, $j_{\pm} = j \pm 1/2$, and $k_{\pm} = k \pm 1/2$. The edge lengths are given by 
\begin{equation}
\Delta z = \frac{2a}{N_\theta}, \quad \Delta l_\theta = \Delta \theta = \frac{1}{\sin\theta} \Delta\tilde{\theta} = \cos^{-1} (\tilde{\theta}_{j_+}) - \cos^{-1} (\tilde{\theta}_{j_-}), \quad \Delta l_\phi = \Delta\phi \sin\theta.
\end{equation}
The volume of each cell is $\Delta V = \Delta z \Delta l_\theta \Delta l_\phi = \Delta z \Delta\tilde{\theta} \Delta \phi$, and the areas of the cell faces are 
\begin{equation}
ABCD \text{ and } EFGH: \Delta l_\theta \Delta l_\phi, \quad ADHE: \Delta z\Delta l_\phi^- \text{ and } BCGF: \Delta z\Delta l_\phi^+,\quad ABFE \text{ and } DCGH: \Delta z \Delta l_\theta.
\end{equation}

By applying the divergence theorem to the Smoluchowski equation, the probability conservation law for a computation cell can be written as
\begin{equation}
\frac{\partial}{\partial t}\int \Psi \,dV + \left(\int_{EFGH} - \int_{ABCD}\right) J_z \,d\theta \,d\phi + \left(\int_{BCGF} - \int_{ADHE}\right) J_\theta \,dz\,d\phi + \left(\int_{DCGH} - \int_{ABFE}\right)J_\phi \,dz\,d\theta = 0.
\end{equation}
In discrete form,
\begin{equation}\label{finite_volume}
\begin{aligned}
\frac{\Psi_{i,j,k}^{n+1}-\Psi_{i,j,k}^{n}}{\Delta t} + \frac{1}{\Delta z}\left[J_z^n (i_+,j,k) - J_z^n (i_-, j, k)\right] + \frac{1}{\Delta \tilde{\theta}}\left[J_\theta^n(i, j_+, k)\sin\theta_{j_+} - J_\theta^n(i, j_-, k)\sin\theta_{j_-}\right] \\
+ \frac{1}{\Delta l_{\phi}} \left[J_{\phi}^n(i,j,k_+) - J_{\phi}^n(i,j,k_-)\right] = 0.
\end{aligned}
\end{equation}
From Eq.~(\ref{flux}), the $z$-component fluxes at the centers of the volume surfaces are 
\begin{equation}
\begin{aligned}
J_z (i_+, j, k) &= \cos\theta_j \frac{\Psi_{i,j,k} + \Psi_{i+1,j,k}}{2} + \frac{U^{\ast}_z (i,j,k) \Psi_{i,j,k} + U^{\ast}_z (i+1,j,k) \Psi_{i+1,j,k}}{2} - D_j \frac{\Psi_{i+1,j,k} - \Psi_{i,j,k}}{\Delta z}, \\
J_z (i_-, j, k) &= \cos\theta_j \frac{\Psi_{i-1,j,k} + \Psi_{i,j,k}}{2} + \frac{U^{\ast}_z (i-1,j,k) \Psi_{i-1,j,k} + U^{\ast}_z (i,j,k) \Psi_{i,j,k}}{2} - D_j \frac{\Psi_{i,j,k} - \Psi_{i-1,j,k}}{\Delta z}.
\end{aligned}
\end{equation}
The $\theta$-component fluxes are
\begin{equation}
\begin{aligned}
J_{\theta}(i, j_+, k) &= \left[\lambda_{j_+} \Psi_{i,j+1,k}\Omega_\phi (i, j+1, k) + (1-\lambda_{j_+}) \Psi_{i,j,k}\Omega_\phi (i, j, k) \right] - D_{\mathrm{r}} \sin\theta_{j_+} \frac{\Psi_{i,j+1,k}-\Psi_{i,j,k}}{\Delta\tilde{\theta}}, \\
J_{\theta}(i, j_-, k) &= \left[\lambda_{j_-}\Psi_{i,j-1,k}\Omega_\phi (i, j-1, k) + (1-\lambda_{j_-}) \Psi_{i,j,k}\Omega_\phi (i, j, k) \right] - D_{\mathrm{r}} \sin\theta_{j_-} \frac{\Psi_{i,j,k}-\Psi_{i,j-1,k}}{\Delta\tilde{\theta}},
\end{aligned}
\end{equation}
where the interpolation weights
\begin{equation}
\begin{aligned}
\lambda_{j_+} &= \dfrac{\cos^{-1} \tilde{\theta}_{j_+} - \cos^{-1} \tilde{\theta}_j}{\cos^{-1}\tilde{\theta}_{j+1} - \cos^{-1}\tilde{\theta}_j},\\
\lambda_{j_-} &= \dfrac{\cos^{-1} \tilde{\theta}_{j_-} - \cos^{-1} \tilde{\theta}_j}{\cos^{-1}\tilde{\theta}_{j-1} - \cos^{-1}\tilde{\theta}_j},
\end{aligned}
\end{equation}
and we have used the conversion from $\theta$ to $\tilde{\theta}$,
\begin{equation}
\left.\frac{d\tilde{\theta}}{d\theta} \right |_{j_{+}} = -\sin\theta_{j_{+}} \Rightarrow \frac{\tilde{\theta}_{j+1} - \tilde{\theta}_{j}}{\theta_{j+1} - \theta_j} = -\sin\theta_{j_+} \Rightarrow \Delta\theta_{j_{+}}= \frac{1}{\sin\theta_{j_+}} \Delta\tilde{\theta}.
\end{equation}
The $\phi$-component fluxes are
\begin{equation}
\begin{aligned}
J_{\phi} (i,j,k_+)  &= -\frac{\Omega_{\theta} (i,j,k)\Psi_{i,j,k} + \Omega_{\theta} (i,j,k+1)\Psi_{i,j,k+1}}{2} - D_{\mathrm{r}}\frac{1}{\sin \theta_j} \frac{\Psi_{i,j,k+1} - \Psi_{i,j,k}}{\Delta\phi}, \\
J_{\phi} (i,j,k_-)  &= -\frac{\Omega_{\theta} (i,j,k-1)\Psi_{i,j,k-1} + \Omega_{\theta} (i,j,k)\Psi_{i,j,k}}{2} - D_{\mathrm{r}}\frac{1}{\sin \theta_j} \frac{\Psi_{i,j,k} - \Psi_{i,j,k-1}}{\Delta\phi}.
\end{aligned}
\end{equation}

For control volumes in the bulk, Eq.~(\ref{finite_volume}) is evolved forward using an explicit scheme. However, special care is required for control volumes located on the boundaries in Eq.~(\ref{theta_range}) and those adjacent to $\theta=0$ and $\theta = \pi$. For boundary volumes, surfaces lying outside the domain do not contribute to the net flux. For more details, see Ref.~\cite{Ezhilan2015}.

\section{Simulation of active Brownian spheroidal particles}
\textit{Translational equation}.---In the collision simulation, bacteria as modeled as spheroid particles. We begin by describing their translational motion. Each particle swims at speed $v_0$ along its orientation vector $\mathbf{p}$, relative to the imposed background flow $\mathbf{u}$. In spherical coordinate, $\mathbf{p} = (\cos\phi\sin\theta, \sin\phi\sin\theta, \cos\theta)$. The translational motion is described by 
\begin{equation}\label{translation}
\boldsymbol{\zeta}\cdot \left(\frac{d\textbf{r}}{dt} - v_0\textbf{p} -\textbf{u}\right) = \boldsymbol{\xi},
\end{equation}
where the resistance tensor is $\boldsymbol{\zeta} = \zeta_{\parallel}\textbf{p}\textbf{p} + \zeta_{\perp}(\textbf{I}-\textbf{p}\textbf{p})$ and $\boldsymbol{\xi}$ is a random force. The parallel and perpendicular components of translational resistance are given by~\cite{Kim2013}
\begin{gather}
\zeta_{\parallel}/(6\pi\mu a) = \frac{8}{3}e^3 \left[-2 e + (1+e^2)l\right]^{-1}, \\
\zeta_{\perp}/(6\pi\mu a) = \frac{16}{3}e^3 \left[2e + (3e^2 - 1)l\right]^{-1},
\end{gather}
where $e$ is the spheroid eccentricity and $l = \ln [(1+e)/(1-e)]$. The random force satisfies the statistics~\cite{Doi2013},
\begin{gather}
\langle \xi_i \rangle = 0,\\
\langle \xi_i \xi_j \rangle = 2k_{\mathrm{B}} T \zeta_{ij};\quad i,j = \{x,y,z\}.
\end{gather}

We multiply both sides of Eq.~(\ref{translation}) by $\boldsymbol{\zeta}^{-1}$ and rescale the random force as $\boldsymbol{\zeta}^{-1}\cdot \boldsymbol{\xi} = M\cdot \boldsymbol{\eta}$, where $\boldsymbol{\eta}$ is a white Gaussian noise with zero mean and unit variance,
\begin{equation}
\langle \eta_i \rangle = 0,\quad \langle \eta_i(t) \eta_j(t') \rangle = \delta_{ij}\delta(t-t').
\end{equation}
The scaling matrix $M$ satisfies
\begin{equation}
M\cdot M^{\mathrm{T}} = 2k_{\mathrm{B}} T \boldsymbol{\zeta}^{-1}.
\end{equation}
The inverse of the resistance matrix can be evaluated as
\begin{equation}
\begin{aligned}
\boldsymbol{\zeta}^{-1} &= (\zeta_{\parallel}^{-1}-\zeta_{\perp}^{-1}) \textbf{p}\textbf{p} + \zeta_{\perp}^{-1}\textbf{I} \\
&= \frac{1}{2}\zeta_{\perp}^{-1}\left[(1 + \gamma_{\perp}) \textbf{I} + (\gamma_{\perp}-1) P(\theta, \phi) \right],
\end{aligned}
\end{equation}
where the drag anisotropy ratio is given by
\begin{equation}
\gamma_{\perp}(\lambda) = \zeta_{\perp}/\zeta_{\parallel} = \frac{2\left[-2 e + (1+e^2)l\right]}{2e + (3e^2 - 1)l},
\end{equation}
and $P(\theta, \phi)$ is a function of  $\phi$ and $\theta$, 
\begin{equation}
P = 2\mathbf{p}\mathbf{p} - \mathbf{I} = \begin{bmatrix}
-1 + 2\cos^2\phi\sin^2\theta & \sin2\phi \sin^2\theta & \cos\phi\sin2\theta \\
\sin2\phi \sin^2\theta & -1 + 2\sin^2\phi\sin^2\theta & \sin\phi\sin2\theta \\
\cos\phi\sin2\theta & \sin\phi\sin2\theta & \cos2\theta
\end{bmatrix}.
\end{equation}
The perpendicular diffusion coefficient is defined as 
\begin{equation}
D_{\perp} = k_{\text{B}}T \zeta_{\perp}^{-1}.
\end{equation}
The equation satisfied by $M$ can then be rewritten as
\begin{equation}
M\cdot M^{\mathrm{T}} = D_{\perp}\left[(1 + \gamma_{\perp}) \textbf{I} + (\gamma_{\perp}-1) P(\theta, \phi)\right].
\end{equation}
Therefore, the translational equation is
\begin{equation}
\frac{d\textbf{r}}{dt} = v_0 \textbf{p} + \textbf{u} + M\cdot \boldsymbol{\eta}.
\end{equation}

\begin{figure}[t]
\centering
\includegraphics[bb= 5 0 350 300, scale=1.0, draft=false]{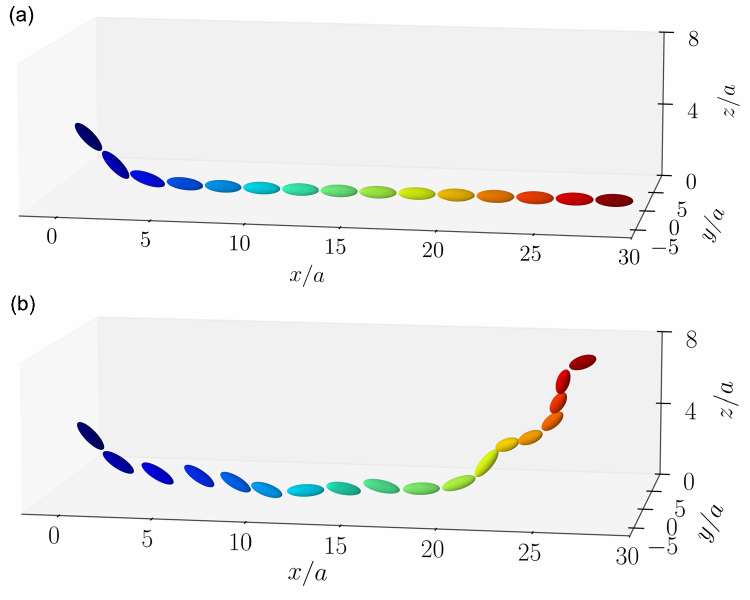}
\caption{Simulated collision process of a swimming spheroidal particle with a nearby no-slip wall (a) without rotational diffusion, (b) with rotational diffusion for $D_r = 0.2$ rad$^2$/s. To demonstrate the effect of steric exclusion enforced by Eq.~(\ref{TR_coupling}) (or equivalently Eq.~(\ref{zero_normal_flux}) in the continuum model), the images flows of the dipole and quadrupole are not included. Time evolves from blue to red. Lengths are scaled by the semi-major axis length $a$. To provide a clear illustration of the collision process with a relatively smooth trajectory, the translational diffusion is switched off in both cases.}
\label{figS3}
\end{figure}
\textit{Rotational equation}.---For the rotational dynamics, we ignore the spinning around the particle's major axis and only consider the rotation around the minor axis. The contribution of the fluid flow to the angular velocity $\boldsymbol{\Omega}$ is given by Jeffrey's equation,
\begin{equation}
\boldsymbol{\Omega} = \frac{1}{2}\nabla \times \textbf{u} + B \textbf{p}\times (E\cdot \textbf{p}),
\end{equation}
where $B = (\beta^2 - 1)/(\beta^2 + 1)$ and the rate of strain tensor $E = [\nabla \textbf{u} + (\nabla \textbf{u})^{\mathrm{T}}]/2$. Therefore, given random rotational noise $\boldsymbol{\eta}^\mathrm{r}$, the angular velocity $\boldsymbol{\omega}$ of the particle can be written as
\begin{equation}\label{angular_velocity}
\boldsymbol{\omega} = \boldsymbol{\Omega} + \sqrt{2D_{\mathrm{r}}}\boldsymbol{\eta}^{\mathrm{r}},
\end{equation}
where the rotational diffusion constant $D_r = k_B T/\zeta_{r}$ with the resistance $\zeta_{r}$ around the minor axis given by
\begin{equation}
\zeta_{\mathrm{r}}/(8\pi\mu a^3) = \frac{4}{3}e^3(2-e^2)\left[-2e + (1+e^2)l\right]^{-1}.
\end{equation}
The random rotational noise $\boldsymbol{\eta}^{\mathrm{r}}$ satisfies the statistics,
\begin{gather}
\langle \eta_i^{\mathrm{r}} \rangle = 0,\\
\langle \eta_i^{\mathrm{r}}(t) \eta_j^{\mathrm{r}}(t') \rangle = \delta_{ij}\delta(t-t'); \quad i,j = \{\theta, \phi\}.
\end{gather}
With the angular velocity known, the particle orientation $\mathbf{p}$ evolves by
\begin{equation}
\frac{d \textbf{p}}{dt} = \left(\boldsymbol{\Omega} + \sqrt{2 D_{\mathrm{r}}}\boldsymbol{\eta}^{\mathrm{r}} \right)\times \textbf{p}.
\end{equation}

\textit{Coupling between translation and rotation.}---When particles are in contact with surfaces, steric exclusion induce a coupling between translational and rotational motions. From Eq.~(\ref{zero_normal_flux}), for $\delta\leq z \leq \delta + a$ (near the bottom plate), the translational velocity component $v_z$ and the angular velocity $\dot{\theta}$ are related by
\begin{equation}\label{TR_coupling}
\begin{dcases}
v_z + a\sin\theta \dot{\theta} = 0, & \text{for }\theta = \cos^{-1}[(z-\delta)/a]\ (<\pi/2),\\
v_z - a\sin\theta\dot{\theta} = 0, & \text{for }\theta = \cos^{-1}[-(z-\delta)/a]\ (>\pi/2).
\end{dcases}
\end{equation}
A similar relation applies near the top plate. Equation~(\ref{TR_coupling}) ensures gradual alignment of the particle with the wall in the absence of rotational diffusion, without explicitly modeling contact force [Fig.~\ref{figS3}(a)]. Rotational diffusion facilitates particle escape from the surface [Fig.~\ref{figS3}(b)]. 

\section{Boundary element method}
\begin{table}[b]
\centering
\begin{tabular}{ c|c|c}
 parameter & symbol & value \\ 
 \hline
 body length & $2 a$ & 2.5 $\mu$m \\
 body width & $2 b$ & 0.9  $\mu$m \\
 motor frequency & $\Omega_{\mathrm{m}}$ & 154 $s^{-1}$ \\
 flagellum axial length & $L$ & 7.0 $\mu$m \\
 flagellum wavelength & $\lambda$ & 2.22 $\mu$m \\
 flagellum helical radius & $R$ & 0.2 $\mu$m \\ 
 flagellum cross-section radius & $\rho$ & 0.012 $\mu$m \\
 fluid viscosity & $\mu$ & $10^{-3}$ Pa$\cdot$s \\
\end{tabular}
\caption{Dimensional physical parameters of bacteria. The no-slip plate is located at $z=0$.}
\label{table1}
\end{table}

\begin{figure}[t]
\centering
\includegraphics[bb= 0 0 365 140, scale=1.0, draft=false]{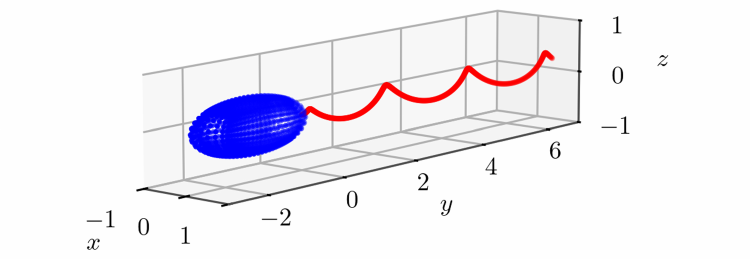}
\caption{A model bacterium used to compute the curvature of swimming trajectory.}
\label{figS4}
\end{figure}

To evaluate how the curvature of the swimming trajectory depends on the plate separation $H$, we simulate the motion of a bacterium propelled by a rotating flagellum using a boundary element method~\cite{Smith2009_2}. The geometrical setup follows Ref.~\cite{Das2018}. The model bacterium consists of a prolate spheroid and a helical flagellum [Fig.~\ref{figS4}], and is driven by a prescribed motor angular velocity $\Omega_{\mathrm{m}}$. The axis of symmetry of the model bacterium is aligned along the $y$-direction. Surface point $(x,y,z)$ on the cell body satisfy,
\begin{equation}\label{cell_body}
\frac{x^2}{b^2} + \frac{y^2}{a^2} + \frac{z^2}{b^2} = 1,
\end{equation}
where $a$ and $b$ are the semi-major and semi-minor axes, respectively, with $a/b > 1$. The flagellum bundle is modeled as a rigid left-handed helix. The helix is parameterized by the axial length $s$ along its axis of symmetry, 
\begin{equation}\label{flagellum}
\begin{aligned}
x &= -E(s) R\sin (k s +\phi), \quad s\in [0, L] \\
y &= s, \\
z &= E(s) R\cos(k s + \phi).
\end{aligned}
\end{equation}
Here, $L$ is the axial length of the flagellum, and the growth function $E(s) = 1- e^{-k^2 s^2}$ controls how quickly the helix reaches its maximum amplitude $R$, which is also the helical radius. The wavenumber is $k$, and wavelength $\lambda = 2\pi/k$. The rotation phase of the helix is controlled by $\phi$. Parameter values are taken from experimental measurements and listed in Table~\ref{table1}.

Denote the surface of the cell body as $S_{\mathrm{b}}$ [Eq.~(\ref{cell_body})] and the centerline of the flagellum as $C_{\mathrm{f}}$ [Eq.~(\ref{flagellum})]. The velocity on the cell body surface and flagellum centerline can be expressed as
\begin{equation}\label{integral_velo}
\mathbf{u}(\mathbf{x}_0) = \frac{1}{8\pi\mu} \iint\limits_{S_{\mathrm{b}}} G (\mathbf{x}_0, \mathbf{x}) \cdot \mathbf{f}_{\mathrm{b}}(\mathbf{x})\,d \mathbf{x} + \frac{1}{8\pi\mu} \int\limits_{C_{\mathrm{f}}} G (\mathbf{x}_0, \mathbf{x}(s)) \cdot \mathbf{f}(\mathbf{x}(s))\,d s, \quad \text{with }\mathbf{x}_0\in S_{\mathrm{b}}, C_{\mathrm{f}}.
\end{equation}
Here, $G(\mathbf{x}_0, \mathbf{x})$ is the Stokeslet solution between two parallel plates~\cite{Liron1976}, the flagellum is parameterized by the axial length $s$, and $\mathbf{f}_{\mathrm{b}}$ and $\mathbf{f}$ represent the force densities on the cell body and flagellum, respectively. The origin is placed at the point where the flagellum  attaches to the cell body.

The kinematic constraint requires that the hydrodynamic velocity $\mathbf{u}(\mathbf{x}_0)$ matches the instantaneous velocity on the cell body and flagellum,
\begin{equation}
\begin{aligned}
\mathbf{u}(\mathbf{x}_0) &= \mathbf{U} + \boldsymbol{\Omega}_\mathrm{b}\times \mathbf{x}_0,\quad \text{with }\mathbf{x}_0\in S_{\mathrm{b}}, \\
\mathbf{u}(\mathbf{x}_0) &= \mathbf{U} + (\boldsymbol{\Omega}_{\mathrm{b}} + \boldsymbol{\Omega}_{\mathrm{m}})\times \mathbf{x}_0, \quad \text{with } \mathbf{x}_0\in C_{\mathrm{f}},
\end{aligned}
\end{equation}
where $\boldsymbol{\Omega}_\mathrm{b}$ is the angular velocity of the cell body and $\boldsymbol{\Omega}_\mathrm{m}$ is the motor velocity (the relative velocity between the flagellum and cell body). For a free-swimming bacterium, the total hydrodynamic force and torque acting on it should be zero, i.e., 
\begin{gather}
\iint\limits_{S_{\mathrm{b}}} \mathbf{f}_{\mathrm{b}} (\mathbf{x})\,d\mathbf{x} + \iint\limits_{C_{\mathrm{f}}} \mathbf{f} (\mathbf{x}(s))\,ds = 0, \label{force-free} \\
\iint\limits_{S_{\mathrm{b}}} \mathbf{x}\times \mathbf{f}_{\mathrm{b}} (\mathbf{x})\,d\mathbf{x} + \int_{C_\mathrm{f}} \mathbf{x}\times \mathbf{f} (\mathbf{x}(s))\, ds = 0. \label{torque_free}
\end{gather}
The torques are computed relative to the attachment point. 

Equations~(\ref{integral_velo})--(\ref{torque_free}) constitute a complete system of linear equations for the unknown force densities $\mathbf{f}$, translational velocity $\mathbf{U}$, and rotational velocity $\boldsymbol{\Omega}$.

\section{Comparison of full bacterial distribution between computations and experiments}

\begin{figure}[htbp]
\centering
\includegraphics[width=0.6\textwidth]{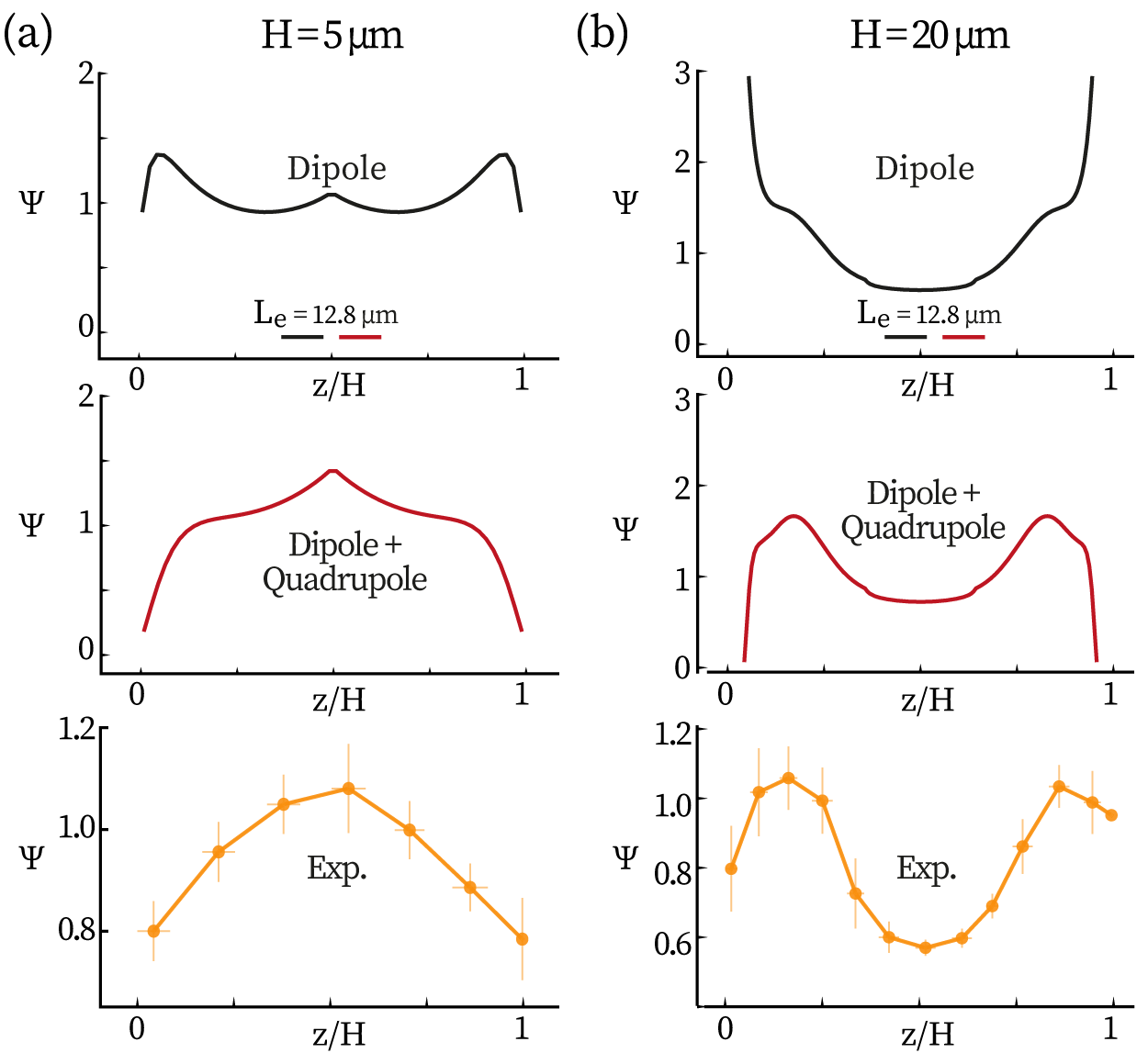}
\caption{Comparison between bacterial distributions that are computed and experimentally measured. Results of solving Smoluchowski equation [Eq. (3) in the main text] with only the dipole term (upper row), with both dipole and quadrupole terms (middle row), and corresponding experimental measurements (lower row), for (a) $H=5$~\um and (b) $H=20$~\um, respectively. Symbols and error bars denote mean $\pm$ SD over 3--4 experiments.}
\label{figS5}
\end{figure}

We compare experimental measurements of bacterial spatial distributions with predictions from the Smoluchowski equation. Figure~\ref{figS5} presents predictions of the Smoluchowski equation, with and without force-quadrupole term, alongside experimental measurements at $H=5$~\um and $H=20$~\um. In both cases --- bulk accumulation [Fig.~\ref{figS5}(a)] and surface accumulation [Fig.~\ref{figS5}(b)] --- including the quadrupole term is essential to reproduce the experimental observations.

\clearpage
\bibliographystyle{apsrev4-2}
\bibliography{reference}